\documentclass[%
 aip,
 jcp,%
 amsmath,amssymb,
reprint,%
]{revtex4-1}

\newcommand{\R}{{\mathbb{R}}}  

\usepackage{graphicx}
\usepackage{dcolumn}
\usepackage{bm}
\RequirePackage[colorlinks=true,linkcolor=blue,,urlcolor=blue,citecolor=blue]{hyperref}

\def\rtx@aipjcp{%
\typeout{Using journal substyle \@journal.}%
\@booleanfalse\authoryear@sw
}

\begin{document}


\title{Chaotic Dynamics in Multidimensional Transition States}           

\author{Ali Allahem }
\email{a.allahem@lboro.ac.uk}
\author{Thomas Bartsch}
\email{t.bartsch@lboro.ac.uk}

\affiliation{ 
Department of Mathematical Science, Loughborough University, LE11 3TU, United Kingdom
}


\date{\today}

\begin{abstract}

The crossing of a transition state in a multidimensional reactive system is mediated by invariant geometric objects in phase space: An invariant hyper-sphere that represents the transition state itself and invariant hyper-cylinders that channel the system towards and away from the transition state. The existence of these structures can only be guaranteed if the invariant hyper-sphere is normally hyperbolic, i.e., the dynamics within the transition state is not too strongly chaotic.  We study the dynamics within the transition state for the hydrogen exchange reaction in three degrees of freedom. As the energy increases, the dynamics within the transition state becomes increasingly chaotic. We find that the transition state first looses and then, surprisingly, regains its normal hyperbolicity. The important phase space structures of transition state theory will therefore exist at most energies above the threshold.

\end{abstract}

\pacs{82.20.Db, 34.50.Lf, 05.45.-a}

\keywords{Transition state theory, NHIM.}

\maketitle 

\section{\label{sec:level1}Introduction}

Transition state theory\cite{Truhlar83,Truhlar96,Miller98,Pollak05} (TST) is the cornerstone of reaction rate theory. It assumes that reactant and product regions in phase space can be separated by a dividing surface that all reactive trajectories must cross once and only once. If this condition is satisfied, TST allows one to calculate the (classical) reaction rate exactly. Otherwise, it provides an upper bound to the rate. For this reason, great effort has been devoted to the construction to a surface that is recrossing free or at least minimizes recrossings. 

Pollak and Pechukas \cite{Pollack78,Pollack79a,Pollack79b,Pollack80,Pollack85} identified the optimal dividing surface in collinear exchange reactions with two degrees of freedom: It is given by the projection into configuration space of an unstable  periodic orbit and is called a periodic orbit dividing surface (PODS). This surface will yield an exact reaction rate as long as there is only a single periodic orbit in the transition region \cite{Pollack79b}. Otherwise, the family of periodic orbits can be used to compute both upper and lower bounds to the reaction rate \cite{Pollack80}. The phase space structures in the transition region that lead to the failure of TST have been studied in detail for a variety of two-dimensional reactive systems (see, e.g., Refs.~\onlinecite{Pechukas82,davis85,davis86,davis87,Tiyapan93,Tiyapan94,Tiyapan95,Burghardt94,Burghardt95}).

In systems with more than two degrees of freedom, a recrossing-free dividing surface was found only much later\cite{Uzer02}. Such a surface exists only in phase space, not in configuration space. It is bounded by a high-dimensional invariant hyper-sphere that plays the role of the periodic orbit in the two-dimensional setting. At sufficiently low energies, this hyper-sphere is a normally hyperbolic invariant manifold (NHIM), i.e., the dynamical expansion and contraction rates transverse to the hyper-sphere are larger then those in directions parallel to it. There are two important consequences of normal hyperbolicity\cite{Fen71,Wigginsbook}: First, the invariant hyper-sphere will persist under perturbations of the dynamical system, for example, changes in energy. Second, the hyper-sphere possesses stable and unstable invariant manifolds. These manifolds separate reactive from non-reactive trajectories in phase space. They act as reaction channels that guide the system from the reactant configuration towards the transition state and on into the product region. Their knowledge allows a detailed description of the reaction dynamics that goes well beyond a rate calculation \cite{Hernandez93,Hernandez94,Waalkens04a,Waalkens04b,Waalkens04,Waalkens05,Gabern05}.

The invariant hyper-sphere and its stable and unstable manifolds will persist as long as the sphere is normally hyperbolic. This can be guaranteed for energies just above the reaction threshold. The reaction dynamics within the transition state region is then accurately described by a harmonic approximation~\cite{Uzer02}. The dynamics within the invariant hyper-sphere is therefore completely regular, and the condition of normal hyperbolicity is satisfied. At higher energies, the dynamics within the sphere will become partially chaotic, and a breakdown of normal hyperbolicity may result. Such a scenario has indeed been described in \cite{Li06a,Li06,Chun} for a model reaction. The authors analyse the dynamics with the help of normal form transformations. Because this procedure does in general not converge, it can become difficult, in particular at higher energies, to distinguish the properties of the underlying dynamical system from artefacts of the normal form. In this paper, we will investigate the dynamics within the transition state of a physical system and present a detailed description of those features that lead to a breakdown and, surprisingly, to a subsequent reestablishment of normal hyperbolicity.

The invariant hyper-sphere that embodies the transition state has customarily\cite{Uzer02,Waalkens04a,Waalkens04b,Waalkens04,Waalkens05,Gabern05} been called ``the NHIM.'' Because we are interested in situations in which the sphere fails to be normally hyperbolic, we will avoid that term and call this object the central sphere.

We study the $\text{H}+\text{H}_2$ exchange reaction that has also served as the prototypical example for the analysis of transition state structures in two degrees of freedom. We will focus on the dynamics within the central sphere, viz., within the centre manifold of the saddle point that mediates the exchange. The centre manifold forms an invariant subsystem with two degrees of freedom. It can be analyzed by means of a Poincar\'e surface of section. 

There are two fundamental periodic orbits within the centre manifold:  a symmetric stretch periodic orbit (SSPO) and a bending periodic orbit (BPO). In addition, a secondary  symmetric stretch periodic orbit (ScPO), that is generated by a bifurcation, plays an important role because it gives rise to a large regular island. These periodic orbits undergo a sequence of bifurcations in which they successively lose and regain stability.

To decide at what energies the central sphere is normally hyperbolic, we compute Lyapunov exponents in directions parallel and perpendicular to the sphere for both periodic and non-periodic orbits. We find that the SSPO is  the only orbit that violates the condition of normal hyperbolicity, and that only in a small energy interval. Nevertheless, the dynamics of the exchange becomes more and more complex as the energy is increased, as is evident already in the collinear subsystem.

The contents of this paper are as follows.  In Sec.~\ref{sec:level2} we give an overview of the low-energy phase space structures near a saddle point. In Sec.~\ref{sec:level3}, we present the $\text{H}+\text{H}_2$ exchange reaction in three degrees of freedom.  In Sec.~\ref{sec:level4} we investigate the dynamics within centre manifold. We describe the fundamental periodic orbits and their bifurcations, and we illustrate the dynamics by means of Poincar\'e surface of section plots. In Sec.~\ref{sec:level5} we compute Lyapunov exponents of trajectories within the centre manifold and identify energy intervals in which it is normally hyperbolic.

\section{\label{sec:level2}The phase space structure near a saddle point}

We now present the phase space structure of a linear Hamiltonian system near an equilibrium point of centre $\times$ centre $\times$...$\times$ saddle type \cite{Uzer02}. Close to the saddle point, the dynamics is well described by the harmonic Hamiltonian 
\begin{equation}
H= \frac{1}{2} \sum_{j=1}^{n} p_{j}^{2} + \frac{1}{2} \sum_{j=1}^{n-1} \omega_{j}^{2} q_{j}^{2} - \frac{\lambda^{2}}{2} q_{n}^{2}.
\label{Hndof}
\end{equation}
The corresponding equations of motion are given by
\begin{equation}
	\dot q_j = \frac{\partial H}{\partial p_j}, \qquad
	\dot p_j = -\frac{\partial H}{\partial q_j},
\end{equation}
or, in terms of the phase space vector $x=(q_1,..., q_n, p_1,..., p_n) \in \R^{2n}$, by
\begin{equation}
\dot{x} = - J \cdot \nabla H,
\label{eqmotion}
\end{equation}
where 
\begin{equation}
J = \begin{pmatrix}
     0_{n} & - I_{n}    \\
      I_{n} & 0_{n}
\end{pmatrix}
\label{Jmat}
\end{equation}
and $\nabla H =(\frac{\partial H}{\partial q_1}, ..., \frac{\partial H}{\partial q_n}, \frac{\partial H}{\partial p_1}, ..., \frac{\partial H}{\partial p_n} )$. They read explicitly
\begin{align}
	\dot q_j &= p_j,& \dot p_j &= -\omega_j^2 q_j \quad\text{for $j=1,\dots,n-1$},
		 \nonumber \\
	\dot q_n &=p_n, & \dot p_n &= \lambda^2 q_n.
	\label{harmEqs}
\end{align}

The eigenvalues of the matrix associated to the linearized Hamiltonian vector field around the saddle point are $\pm \lambda$ and $\pm i \omega_j$ where $j=1, ..., n-1$.  The pair real eigenvalues $\pm \lambda$ describe the hyperbolic direction of the saddle point while the complex eigenvalues describe the elliptic directions of the saddle point, i.e., oscillations transverse to the reaction coordinate.

The dynamics described by the Hamiltonian~\eqref{Hndof} has a stationary point at $p_i=q_i=0$ at energy zero. We will study the dynamics at a fixed energy $h>0$ above the reaction threshold. The energy surface is  $(2n-1)$ dimensional and is given by
 \begin{equation}
 \frac{1}{2} \sum_{i=1}^{n} p_{i}^{2} + \frac{1}{2} \sum_{i=1}^{n-1} \omega_{i}^{2} q_{i}^{2} - \frac{\lambda^{2}}{2} q_{n}^{2} = h > 0.
 \label{Endof}
\end{equation}
From (\ref{Endof}), we can see the that the section through the energy surface at fixed $q_n$ is a $(2n-2)$ sphere with radius $\sqrt{h + \frac{\lambda^{2}}{2} q_{n}^{2}}$. Thus the energy surface is a hyper-cylinder  $S^{2n-2} \times \R$.

The centre manifold of the equilibrium point contains all trajectories that remain trapped close to the equilibrium for all time in the infinite future and the infinite past. It is given by $q_{n}=p_{n} = 0$. This surface is invariant because the equations of motion~\eqref{harmEqs} imply $\dot q_n=\dot p_n=0$. It has dimension $2n-2$.  For a fixed energy $h>0$ it intersects the energy shell in a surface that satisfies
 \begin{equation}
 \frac{1}{2} \sum_{i=1}^{n-1} p_{i}^{2} + \frac{1}{2} \sum_{i=1}^{n-1} \omega_{i}^{2} q_{i}^{2}  = h.
 \label{NHIMndof}
\end{equation}
This equation describes an $(2n-3)$ dimensional hyper-sphere $S_{h}^{2n-3}$. This is the central sphere (called the NHIM in earlier studies) that forms the bottleneck for phase space transport from reactants to products. It has stable and unstable manifolds attached to it. These are $(2n-2)$ dimensional manifolds, denoted by $W^{s}(S_{h}^{2n-3})$ and $W^{u}(S_{h}^{2n-3})$, respectively. They are given by
\begin{equation}
\begin{split}
 W^{s}(S_{h}^{2n-3}) : \quad&  \frac{1}{2} \sum_{i=1}^{n-1} p_{i}^{2} + \frac{1}{2} \sum_{i=1}^{n-1} \omega_{i}^{2} q_{i}^{2}  = h, \quad p_{n} = - \lambda q_{n}, \\
 W^{u}(S_{h}^{2n-3}) : \quad& \frac{1}{2} \sum_{i=1}^{n-1} p_{i}^{2} + \frac{1}{2} \sum_{i=1}^{n-1} \omega_{i}^{2} q_{i}^{2}  = h, \quad p_{n} =  \lambda q_{n}. \\
 \end{split}
 \label{sunndof}
\end{equation}
These manifolds are referred to as reaction cylinders. Their structure is $S^{2n-3} \times \R$. The stable and unstable manifolds have two branches, forward and backward cylinders, denoted by  $W_f^{s,u}$ and $W_b^{s,u}$, respectively.   
\begin{equation}
\begin{split}
W_{f}^{s}(S_{h}^{2n-3}): \quad  \frac{1}{2} \sum_{i=1}^{n-1} p_{i}^{2} + \frac{1}{2} \sum_{i=1}^{n-1} \omega_{i}^{2} q_{i}^{2}  = h, \quad p_{n} = - \lambda q_{n} >0,\\
W_{b}^{s}(S_{h}^{2n-3}): \quad  \frac{1}{2} \sum_{i=1}^{n-1} p_{i}^{2} + \frac{1}{2} \sum_{i=1}^{n-1} \omega_{i}^{2} q_{i}^{2}  = h, \quad p_{n} = - \lambda q_{n} <0,\\
W_{f}^{u}(S_{h}^{2n-3}): \quad \frac{1}{2} \sum_{i=1}^{n-1} p_{i}^{2} + \frac{1}{2} \sum_{i=1}^{n-1} \omega_{i}^{2} q_{i}^{2}  = h, \quad p_{n} =  \lambda q_{n} >0 ,\\
W_{b}^{u}(S_{h}^{2n-3}): \quad \frac{1}{2} \sum_{i=1}^{n-1} p_{i}^{2} + \frac{1}{2} \sum_{i=1}^{n-1} \omega_{i}^{2} q_{i}^{2}  = h, \quad p_{n} =  \lambda q_{n} <0.
 \end{split}
 \label{bfndof}
\end{equation}

A recrossing-free dividing surface  is a $(2n-2)$ dimensional hyper-sphere that is defined by setting $q_n =0$. It has codimension one in the energy shell and separates reactant from product regions. Each trajectory that crosses the dividing surfaces crosses only once, going from reactants to products if $p_n>0$ or from products to reactants if $p_n<0$. The only exceptions are trajectories within the central sphere, which have $p_n=0$ and remain in the dividing surface for all times. The central sphere is an equator of the dividing surface and splits the dividing surface into two hemispheres with $p_n>0$ and $p_n<0$ that mediate forward and backward reactions, respectively.

The internal dynamics of the central sphere, according to~\eqref{harmEqs}, is described in this approximation by a multidimensional harmonic oscillator and is therefore completely regular. In the transverse direction the central sphere is unstable because it is balanced near the top of an energetic barrier. It is therefore normally hyperbolic. This feature guarantees that both the central sphere and its stable and unstable manifolds persist in the full anharmonic system, at least at low energies where the central sphere lies close to the equilibrium point and the anharmonic terms are small. At higher energies, the invariant manifolds, assuming they exist, can be approximated via normal form transformations\cite{Uzer02,Waalkens04b,Waalkens04,Li06,Li06a}. In the present work, we will avoid normal forms and investigate the persistence of the invariant manifolds by direct numerical simulation.

\section{\label{sec:level3}The hydrogen exchange reaction}

The hydrogen exchange reaction $\text{H} + \text{H}_{2} \rightarrow \text{H}_{2} + \text{H}$ involves  three  atoms. Consequently, if the atoms are assumed to move in three-dimensional space, the reaction is described by nine degrees of freedom. Three of these, which represent the centre of mass motion,  can be separated directly. Of the remaining six degrees of freedom, three describe spatial rotations of the complex and three describe vibrations. However, the attempt to separate rotational from vibrational degrees of freedom leads to a vibrational phase space that is singular for all collinear configurations, which are invariant under rotations around the axis on which the atoms lie\cite{Littlejohn97,Littlejohn98a}.

\begin{figure}[t]
\includegraphics[width=3.2in]{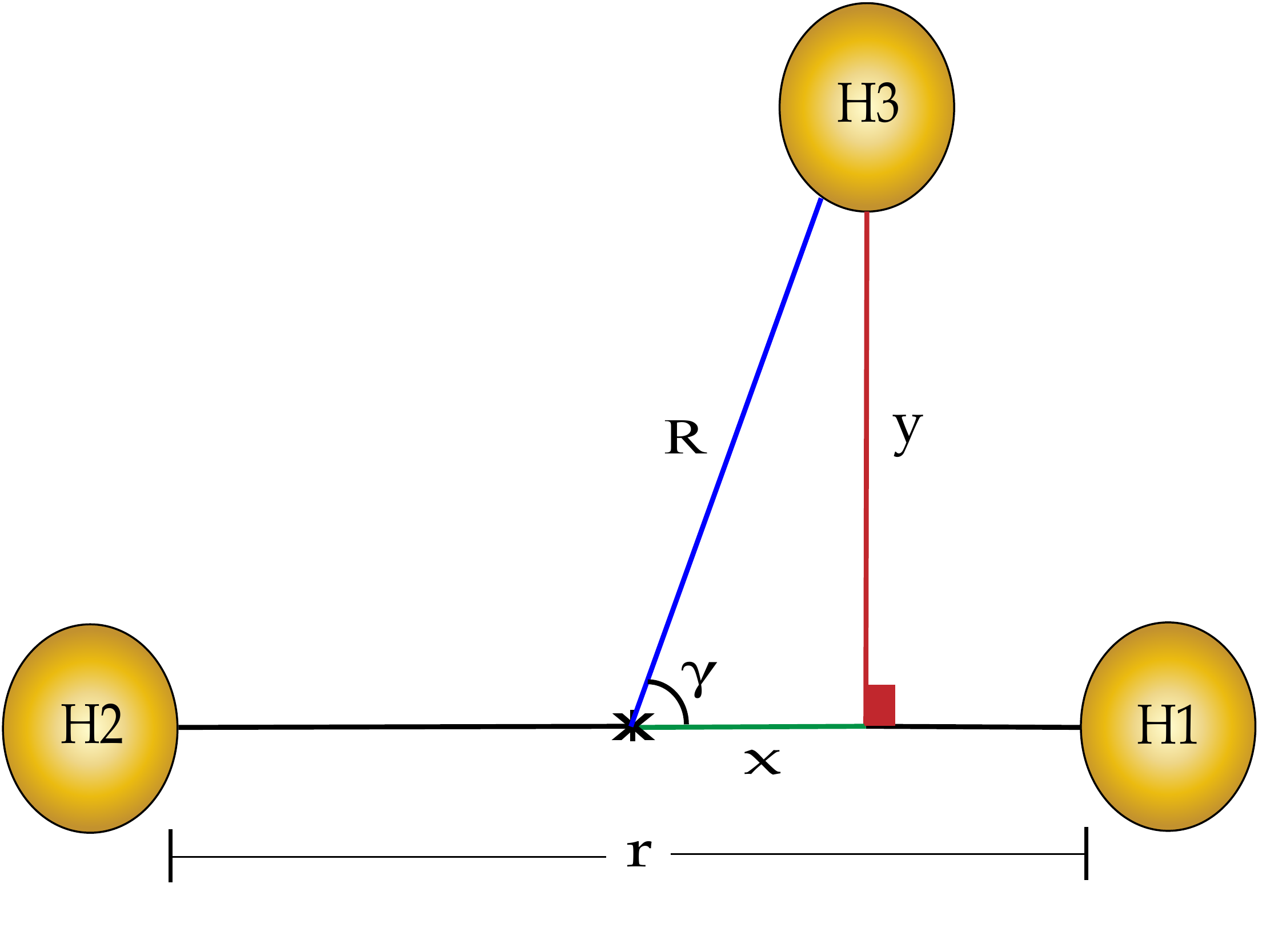}
 \caption{\label{fig:1}The coordinates for $H + H_{2}$ exchange reaction.}
 \end{figure}

The origin of this singularity can be illustrated with the help of Fig.~\ref{fig:1}. To obtain the configuration space of the vibrational dynamics, we have to identify all configurations of the reactive complex that can be transformed into each other by translations or rigid rotations. The shape of the complex can then be described by the three coordinates $r$, $x$ and $y$, where $r$ is the distance between H1 and H2, $y$ is the perpendicular distance from H3 to the distance $r$, and $x$ is the distance from the midpoint of H1 and H2 to the end of the line through H3 perpendicular to $r$. However, not all those configurations are different: $(r,x,y)$ can be transformed $(r,x,-y)$ by a rotation around the axis through H1 and H2. To resolve this ambiguity, the configuration space must be restricted to the half space $(r,x,y\ge 0)$. It has a boundary that is formed by the collinear configurations with $y=0$.  The dynamics must necessarily be singular at these configurations.

We are mainly interested in studying the dynamics in the vicinity of the saddle point that marks the transition region for the exchange reaction. Unfortunately, the activated complex is collinear at the saddle point, and the ensuing singularity makes it difficult to analyze the dynamics. To circumvent this difficulty, we regard configurations with positive and negative values of $y$ as different. This convention, which has also been employed in previous studies~\cite{Waalkens04,Jaffe05}, can physically be interpreted as constraining the three atoms to move in a plane. The full system then has six degrees of freedom, two of which correspond to the centre of mass motion and one to planar rotations. The remaining three degrees of freedom, which can be described, for example, by the three coordinates $r$, $x$ and $y$ of Fig.~\ref{fig:1}, describe the vibrational dynamics of the complex. Because the collinear configurations are not invariant under rotations in the plane, or equivalently, because configurations with positive and negative $y$ cannot be transformed into each other through planar rotations, the symmetry-reduced phase space does not have singularities. It is well suited to an investigation of the dynamics near the saddle point. 

We will study the vibrational dynamics of the complex at zero angular momentum. The Hamiltonian is then given by 
\begin{equation}
H = \frac{1}{m_{H}}\left[ p_{r}^{2} + \tfrac{3}{4} (p_{x}^{2}+p_{y}^{2}) + \frac{(x p_y - y p_x)^2}{r^{2}} \right] + V( r, x, y),
\label{newHamil}
\end{equation}
where $m_H$ is the mass of the hydrogen atom ($m_H$ =  1.00794 amu). The expression for the kinetic energy is derived  from that given, for example, by Waalkens \textit{et al}~\cite{Waalkens04}. This transformation is shown in the appendix. We use the potential energy surface $V( r, x, y)$ derived by Porter and Karplus \cite{Porter63}. Distances will be measured in atomic units (a.u.) and energies in electron volts (eV), with the potential energy of three isolated hydrogen atoms chosen as zero.

The potential energy $V(r,x,y)$ has two reflection symmetries: $x\mapsto -x$ and $y\mapsto -y$. That these transformations must leave the potential invariant is clear from Fig.~\ref{fig:1} because the three atoms are identical. The two reflections of configuration space are extended to phase space by the canonical transformations
\begin{align}
	\label{reflections}
	P_x:\quad & (r, x, y, p_r, p_x, p_y) \mapsto (r, -x, y, p_r, -p_x,  p_y), \nonumber \\
	P_y:\quad & (r, x, y, p_r, p_x, p_y) \mapsto (r, x, -y, p_r, p_x,  -p_y).
\end{align}
Both of these are symmetries of the Hamiltonian~\eqref{newHamil}, as is their composition
\begin{align*}
	P_x\circ P_y:\quad & (r, x, y, p_r, p_x, p_y) \mapsto (r, -x, -y, p_r, -p_x,  -p_y).
\end{align*}

Corresponding to the two reflection symmetries~\eqref{reflections} there are two subsystems with two degrees of freedom. They contain all configurations that are invariant under one of the reflections.

The reactive complex is invariant under $P_y$ if $y=p_y=0$. These are precisely the collinear configurations. The Hamiltonian of the collinear case is given by
\begin{equation}
H = \frac{1}{m_{H}}\left[ p_{r}^{2} + \tfrac{3}{4} p_{x}^{2}  \right] + V( r, x).
\label{collinearHamil2dof}
\end{equation}
Numerous researchers including Pollak and co-workers~\cite{Pollack78, Pollack79a, Pollack79b, Pollack80} and most recently I\~{n}arrea \textit{et al} \cite{Inarrea11} have studied the collinear hydrogen exchange reaction.

The subsystem invariant under the reflection $P_x$ contains all axially symmetric configurations with $x=p_x=0$. The dynamics within this subsystem is described by the Hamiltonian
\begin{equation}
H = \frac{1}{m_{H}}\left[ p_{r}^{2} + \tfrac{3}{4} p_{y}^{2}  \right] + V(r, y).
\label{newHamil2dof}
\end{equation}

The saddle point of the Porter-Karplus potential energy surface is located at the symmetric collinear configuration $(r, x,y) = (r_\text{S}, \;0,\;0)$ with $r_\text{S} = 3.40166\,\text{a.u.}$ To obtain a harmonic approximation of the dynamics close to the saddle point, we expand the Hamiltonian~\eqref{newHamil} in a Taylor series up to second order. The last term in the kinetic energy will not contribute because it is of fourth order. Due to its symmetries the expansion of the potential energy cannot contain any terms of odd order in either $x$ or $y$. Up to an additive constant the harmonic Hamiltonian must therefore be of the form
\begin{equation}
	\label{harmHamil}
	H_2  = \frac{1}{m_{H}}\left[ p_{r}^{2} + \tfrac{3}{4} (p_{x}^{2}+p_{y}^{2}) \right] 
		+ a (r-r_\text{S})^2 + b x^2 + c y^2
\end{equation}
with constants $a$, $b$, $c$ that cannot be determined from symmetry considerations. Thus, the dynamics in $r$, $x$ and $y$ will decouple in the harmonic approximation. Because the expansion point is a saddle, the dynamics must be unstable in one of the three coordinates, namely the reaction coordinate. In the reactant and product states the middle atom H3 is bound to either H1 or H2, whereas the third atom is far away. It is therefore plausible to identify the reaction coordinate with the coordinate $x$ that brings H3 closer to one or the other atom. Indeed, the expansion of the Porter-Karplus potential shows that the coefficient $b$ is negative whereas $a$ and $c$ are positive.

As a consequence, the symmetric subsystem $x=p_x=0$ in which the motion in the reaction coordinate is suppressed forms the centre manifold of the transition state, i.e. it contains all configurations  in which the system oscillates around the unstable equilibrium point. The symmetry of the system makes it easy to identify the centre manifold without laborious calculations. It allows us to avoid the normal form calculations that are required in reactive systems without this symmetry.\cite{Uzer02,Li06,Li06a} 

\section{\label{sec:level4}Dynamics within the centre manifold}

The central sphere that controls transport through the transition state at low energies can be identified with the energy shell within the centre manifold, as described in Sec.~\ref{sec:level2}. As we aim to investigate the breakdown of the low-energy phase space structures, we will start by studying the dynamics within the centre manifold.

\begin{figure}[t]
\includegraphics[width=3.5in]{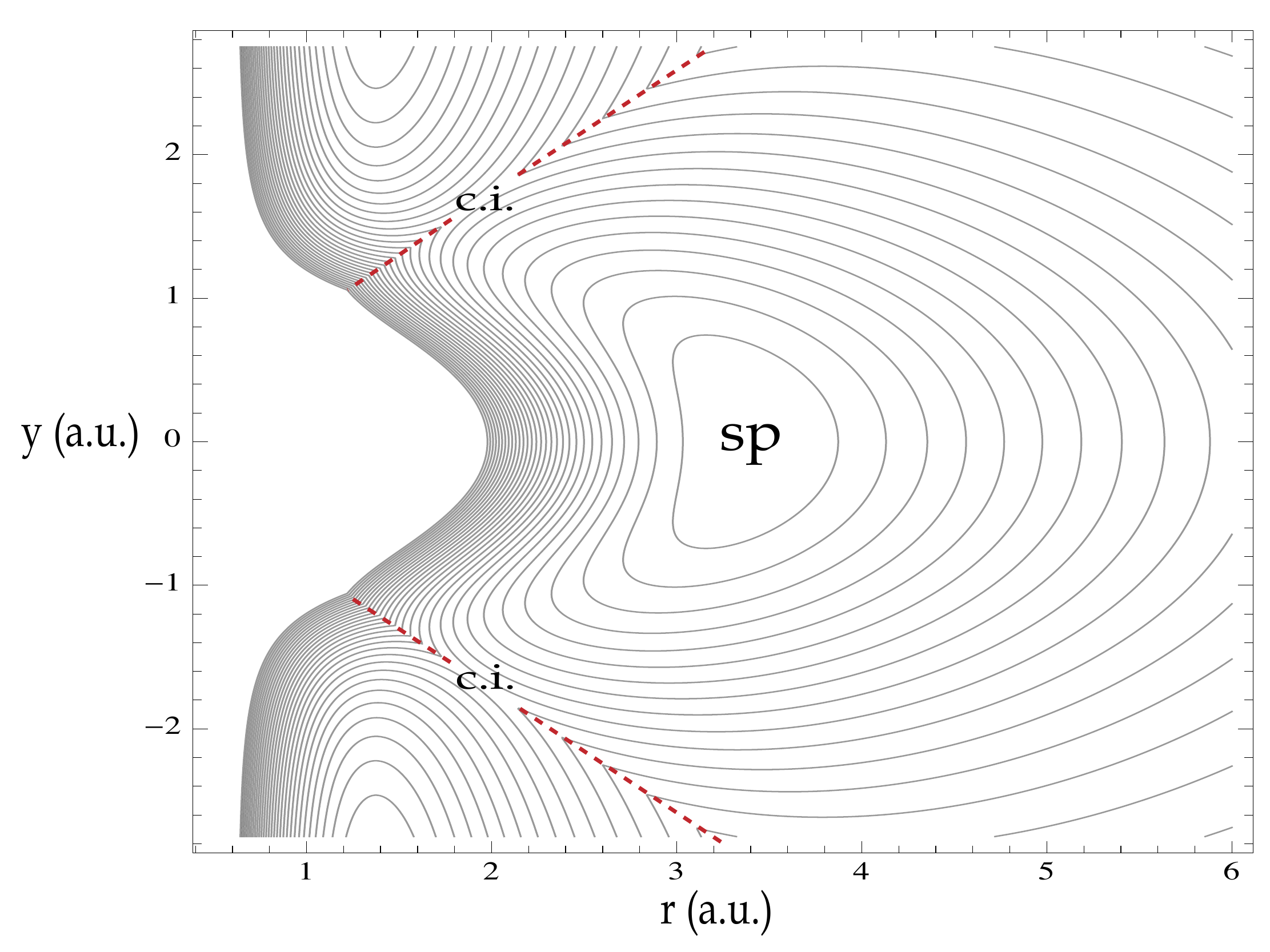}
 \caption{\label{fig:2}The contour plot of the potential energy surface of the centre manifold ($x=0$) of the equilibrium saddle point of the Hamiltonian flow. (sp) refers to saddle point and (c.i.) refers to the conical intersection.}
 \end{figure}

Fig. \ref{fig:2} shows a contour plot of the potential energy for symmetric configurations (i.e. within the centre manifold, with $x=0$). The saddle point (sp) of the three-dimensional system appears as a minimum. It lies at (3.40166 a.u., 0)  in $r, y$ coordinates and has energy value  $-4.3504$ eV.  The second prominent feature of the potential is a conical intersection ridge. It occurs at equilateral configurations, where  $y = \frac{\sqrt{3}}{2} r$. For these configurations the too lowest electronic states are degenerate. As a consequence, the potential energy surface, which gives the energy of the lowest state, is not smooth at the intersection. The lowest point on the ridge occurs at  $r=1.90352 $ a.u., $y = \pm 1.64849$ a.u. with energy $E_{c.i.}=-1.9514$ eV. Above this energy, a new reaction channel opens in which the central atom ($H3$ in Fig. \ref{fig:1}) can escape across the ridge, leaving the two outer atoms bound as a molecule. The transition across the conical intersection cannot be described by classical mechanics. We will restrict our following investigations to energies below $E_{c.i.}$. As we will see, complicated dynamics develop well below this threshold. 

For energies between the saddle point and the conical intersection, the contour line of the potential energy is topologically a circle. The energy shell in phase space (within the centre manifold) has therefore the same topology as it has in the harmonic approximation, i.e., it is a three-dimensional hyper-sphere that we have called the central sphere. At low energies, it is normally hyperbolic. As the energy increases, normal hyperbolicity might, and indeed will, be destroyed. We know from these simple considerations, however, that the central sphere will persist even at energies where it is not normally hyperbolic.

\begin{figure*}
\includegraphics[width=5.5in]{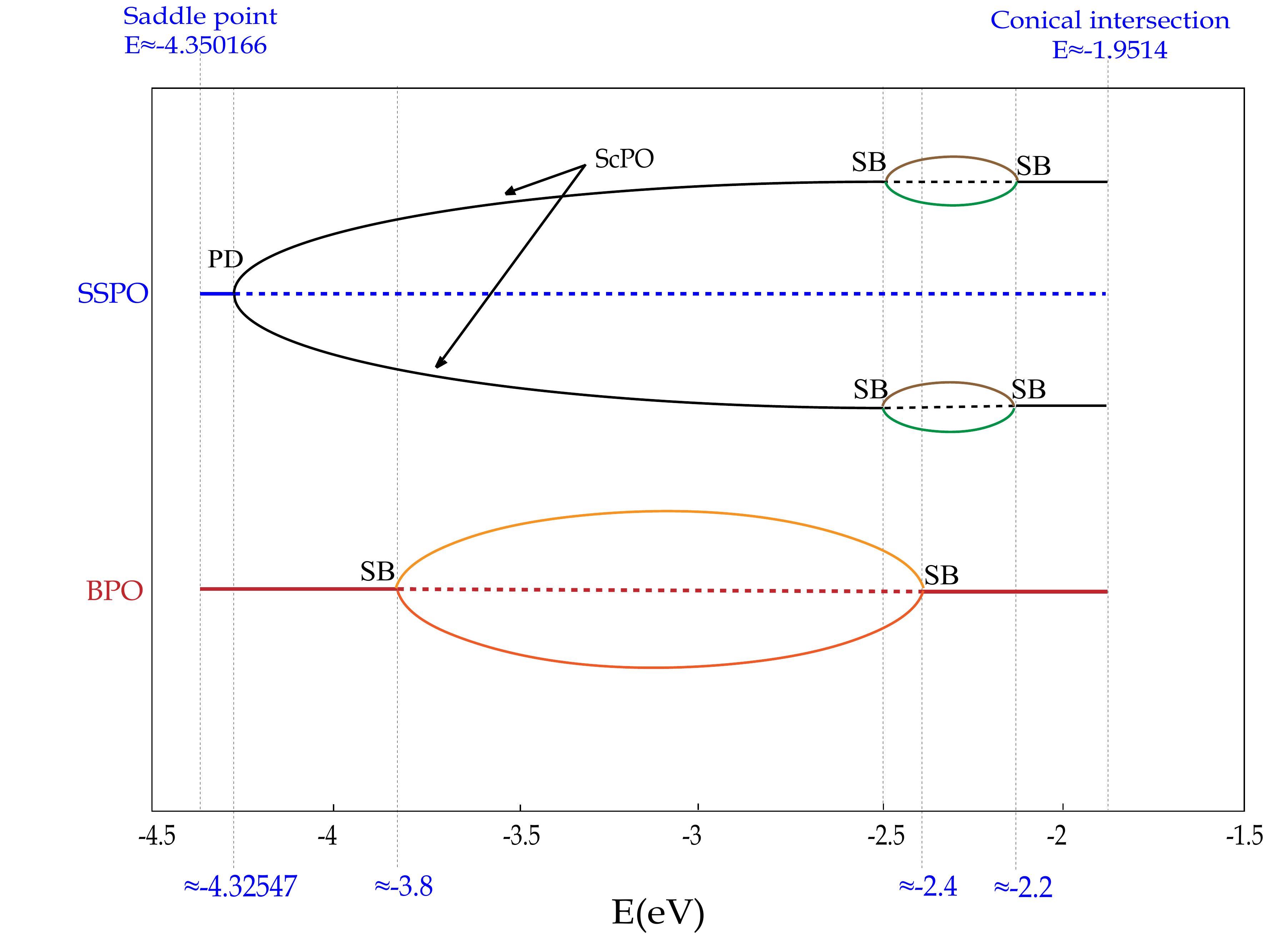}
\caption{\label{fig:4}Bifurcation diagram for the system within centre manifold. The solid and dashed black lines denote stable and unstable periodic orbits, respectively. PD refers to periodic doubling bifurcation and SB refers to symmetry breaking bifurcation.}
\end{figure*}

At energies close to the saddle point where the harmonic approximation is accurate, the dynamics within the central manifold can be described by two normal mode vibrations, a symmetric stretch and a bend of the activated complex. Their frequencies can be obtained from a second-order Taylor series expansion of the potential, i.e., from the constants $a$ and $c$ in Eq.~\eqref{harmHamil}, as $\omega_{SSPO} = 4.1121 \times 10^{14}$ s$^{-1}$ and $\omega_{BPO} =  1.8458 \times 10^{14}$ s$^{-1}$. Both normal mode periodic orbits are stable with respect to a perturbation of initial conditions within the centre manifold. As the energy increases, they undergo a sequence of bifurcations in which they lose their stability and give rise to further stable periodic orbits, as illustrated schematically in Fig. \ref{fig:4}.

Fig.~\ref{fig:4} shows the bifurcation diagram of the two fundamental periodic orbits, the symmetric stretch (SSPO) and the bend (BPO) within the centre manifold.  The first bifurcation occurs in the SSPO at $E \approx -4.32547$ eV, just above the saddle point energy. The SSPO undergoes a period doubling bifurcation: It becomes unstable and a new stable periodic orbit with twice the period appears. We will see that this periodic orbit plays an important role in structuring the dynamics within the centre manifold. We will call it the secondary symmetric stretch periodic orbit (ScPO).

\begin{figure}[t]
\includegraphics[width=3.3in]{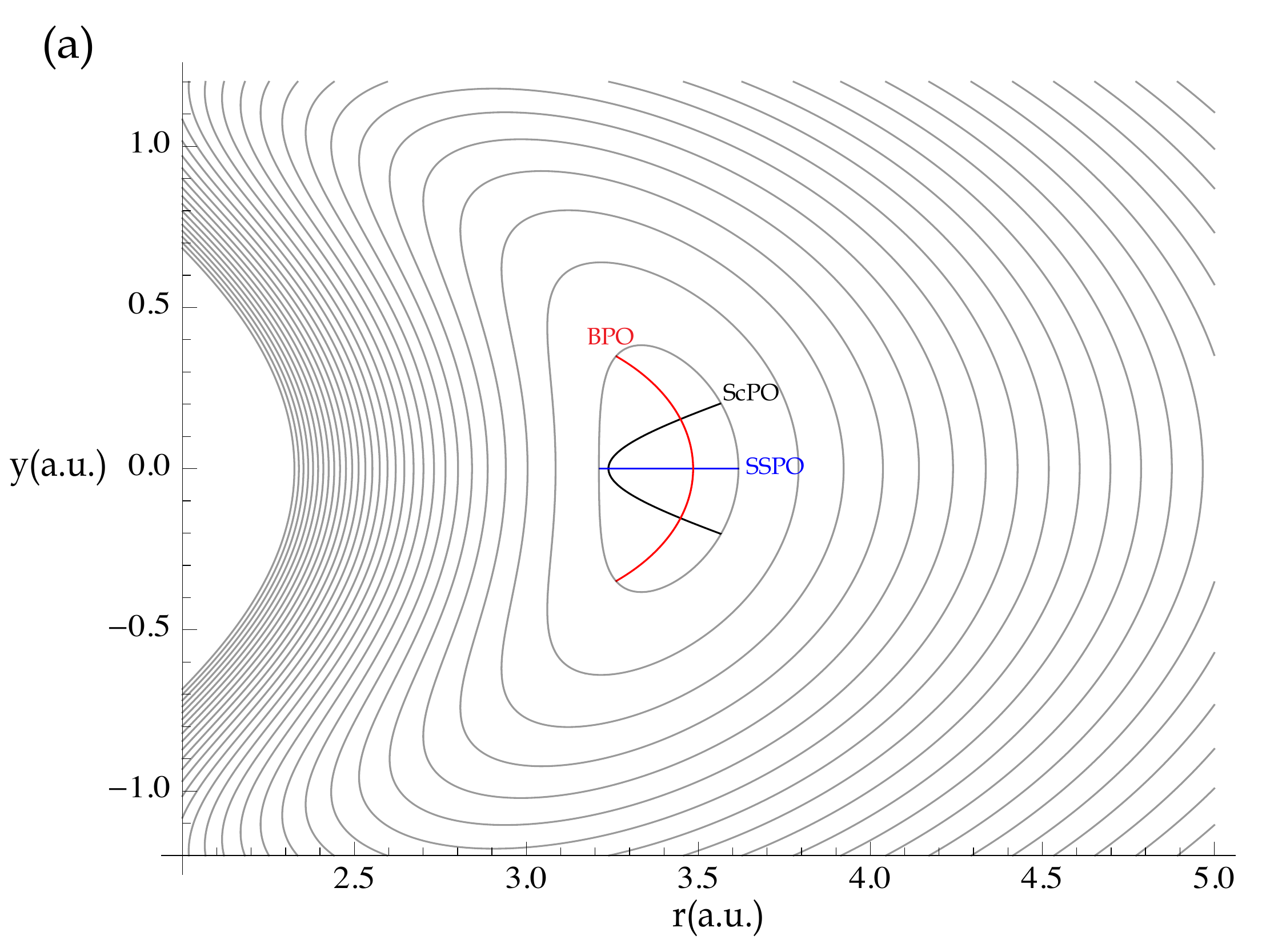}
\includegraphics[width=3.3in]{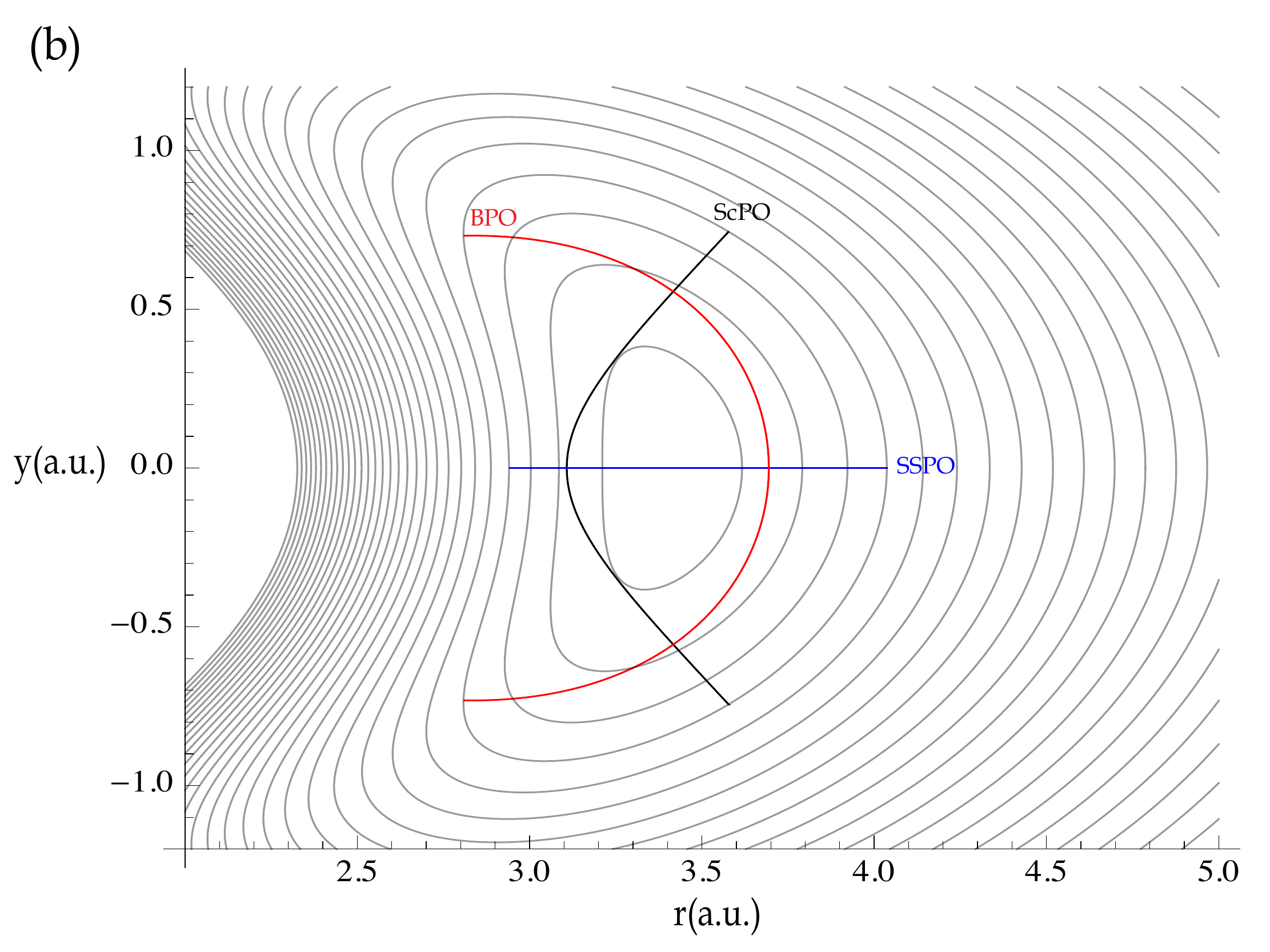}
 \caption{\label{fig:5}Periodic orbits on the potential energy of centre manifold at energies (a) $E=-4.3$ eV, (b) $E=-4.0$ eV.}
 \end{figure}

The configuration space projections of the fundamental periodic orbits are shown in Fig.~\ref{fig:5} for energies $E = -4.3$ eV and $E = -4.0$ eV. It can be clearly seen that even though the ScPO is generated by a bifurcation from the SSPO, it takes on pronounced bending character at higher energies. All three periodic orbits are invariant under the reflection $P_y$. The SSPO is located within the collinear subsystem, which means that each point on the SSPO is invariant under reflection. This is not true for the ScPO and BPO. These periodic orbits are invariant in the sense that any point on one of these orbits is mapped under reflection to a different point on the same orbit. Periodic orbits of this type can undergo symmetry breaking bifurcations that do not exist in systems without reflection symmetries\cite{DeAguiar87,DeAguiar88}: A stable periodic orbit that is invariant under reflection turns unstable and gives rise to two stable periodic orbits that are not invariant, but are mirror images of each other. The asymmetric periodic orbits have roughly the same period as the symmetric one.

\begin{figure}[t]
\includegraphics[width=3.2in]{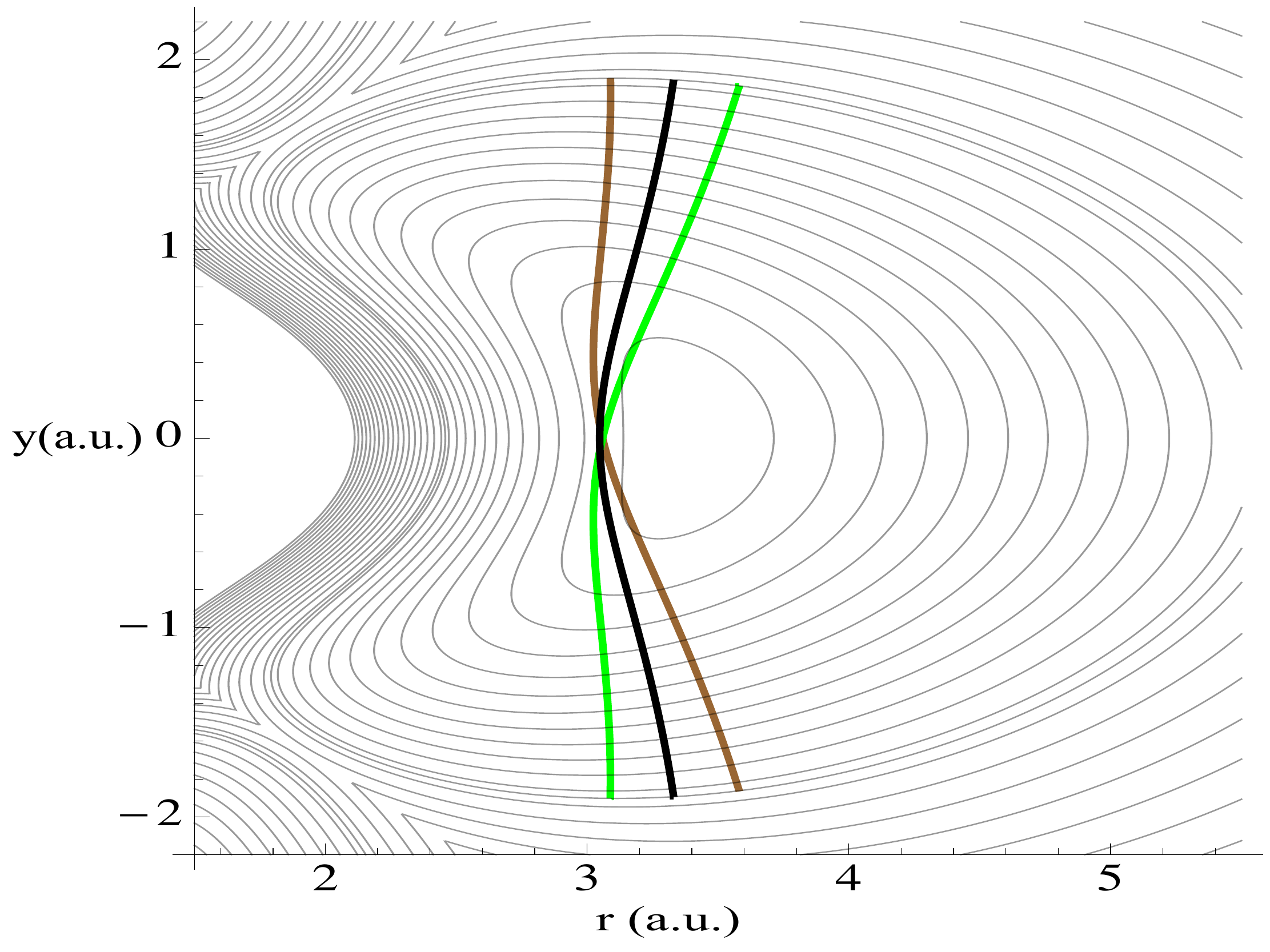}
 \caption{\label{fig:42}The ScPO (black) and its branches (green and brown) at $E = -2.3$~eV.}
 \end{figure}

A symmetry breaking bifurcation of the ScPO occurs at the energy $E \approx -2.5$~eV. Fig.~\ref{fig:42} shows the configuration space projections of the ScPO and the two new periodic orbits for energy $E =-2.3$~eV. It is obvious from the figure that the satellite orbits have lost their reflection symmetry. At a higher energy  $E \approx -2.2$ eV  the asymmetric periodic orbits collapse onto the ScPO again and the ScPO regains stability in an inverse symmetry breaking bifurcation.

\begin{figure*}
\includegraphics[width=5.5in]{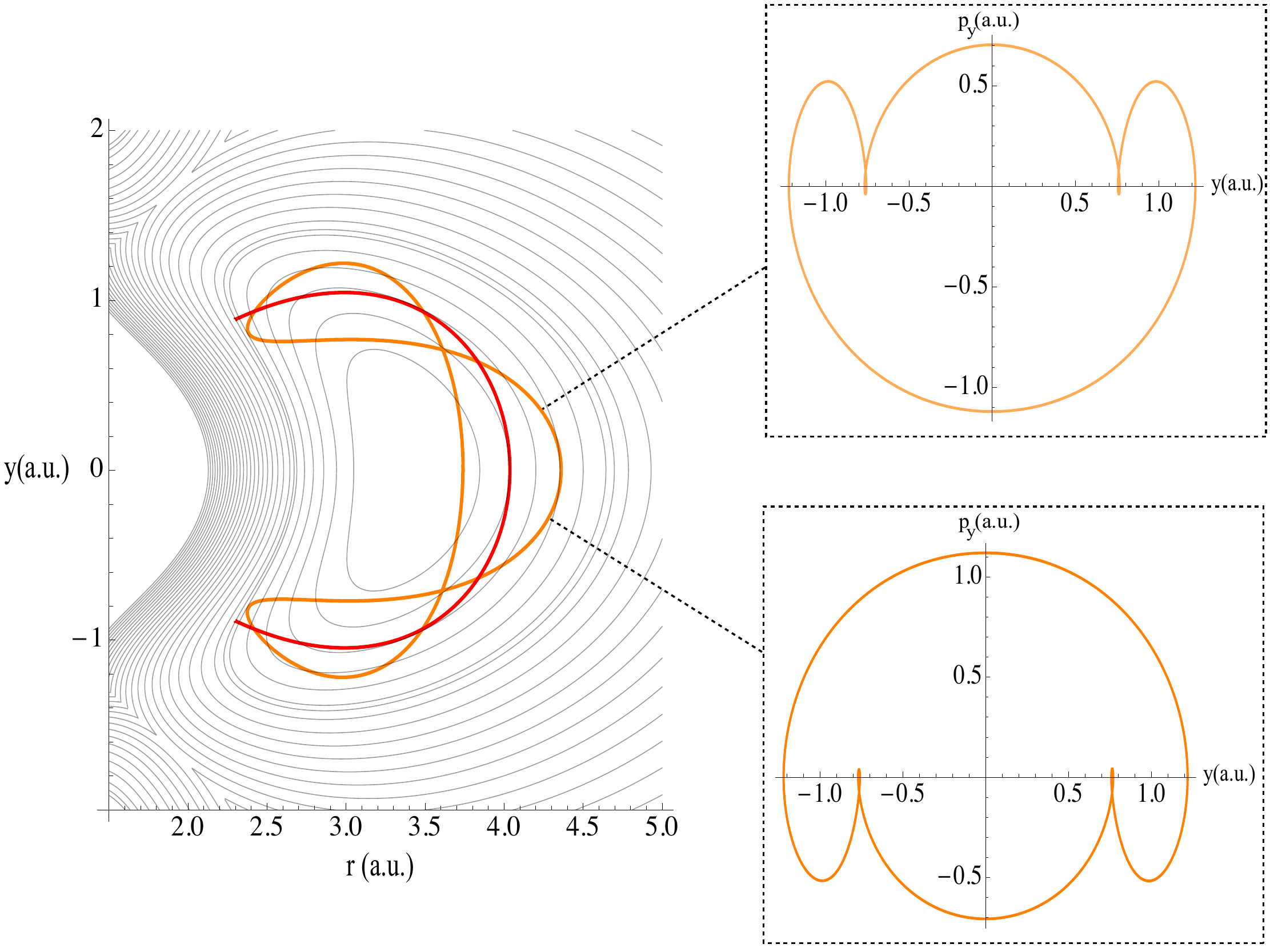}
\caption{\label{fig:41}The BPO and its branches at $E = -3.3$\,eV.}
\end{figure*}

In a similar scenario, the BPO undergoes a symmetry breaking birucation at $E\approx -3.8$~eV, and the two asymmetric periodic orbits thus generated collapse onto the BPO again and disappear at $E\approx -2.4$~eV in an inverse symmetry breaking bifurcation. These three periodic orbits are shown in Fig.~\ref{fig:41}. The two asymmetric orbits have the same projection into configuration space, but, as the phase space figures show, they are traversed in different directions.

\begin{figure*}
\includegraphics[width=6in]{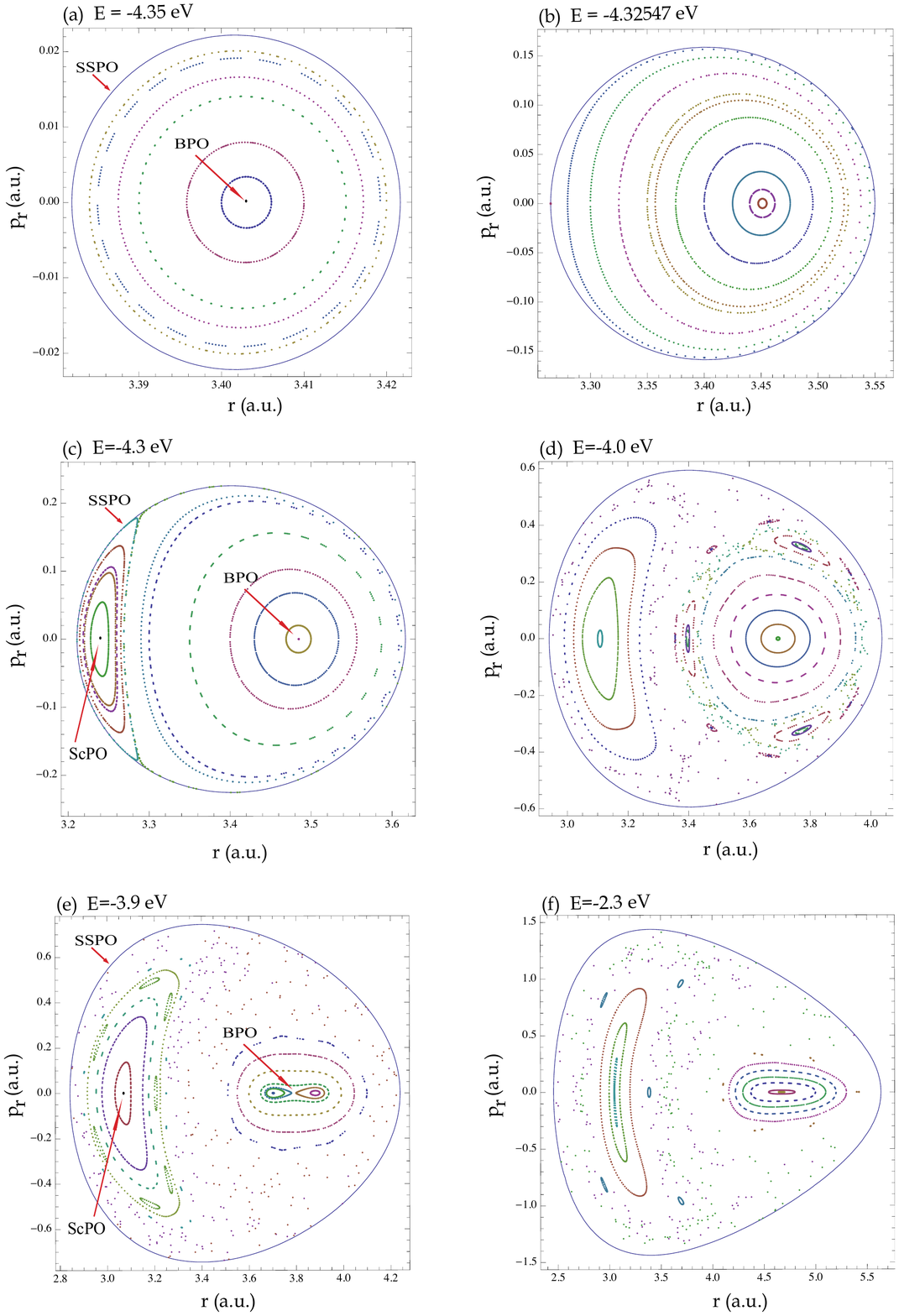}                 
\caption{\label{fig:3}Poincar\'e surface of section $y=0, p_y>0$ for dynamics within the centre manifold at the energies (a) $-4.35$\,eV, (b) $-4.32547$\,eV, (c) $-4.3$\,eV, (d) $-4.0$\,eV, (e) $-3.9$\,eV (BPO is unstable) and (f) $-2.3$\,eV (ScPO is unstable). The main periodic orbits are labeled in (a), (c) and (e).}
\end{figure*}
 
\begin{figure}[t]
\includegraphics[width=3.2in]{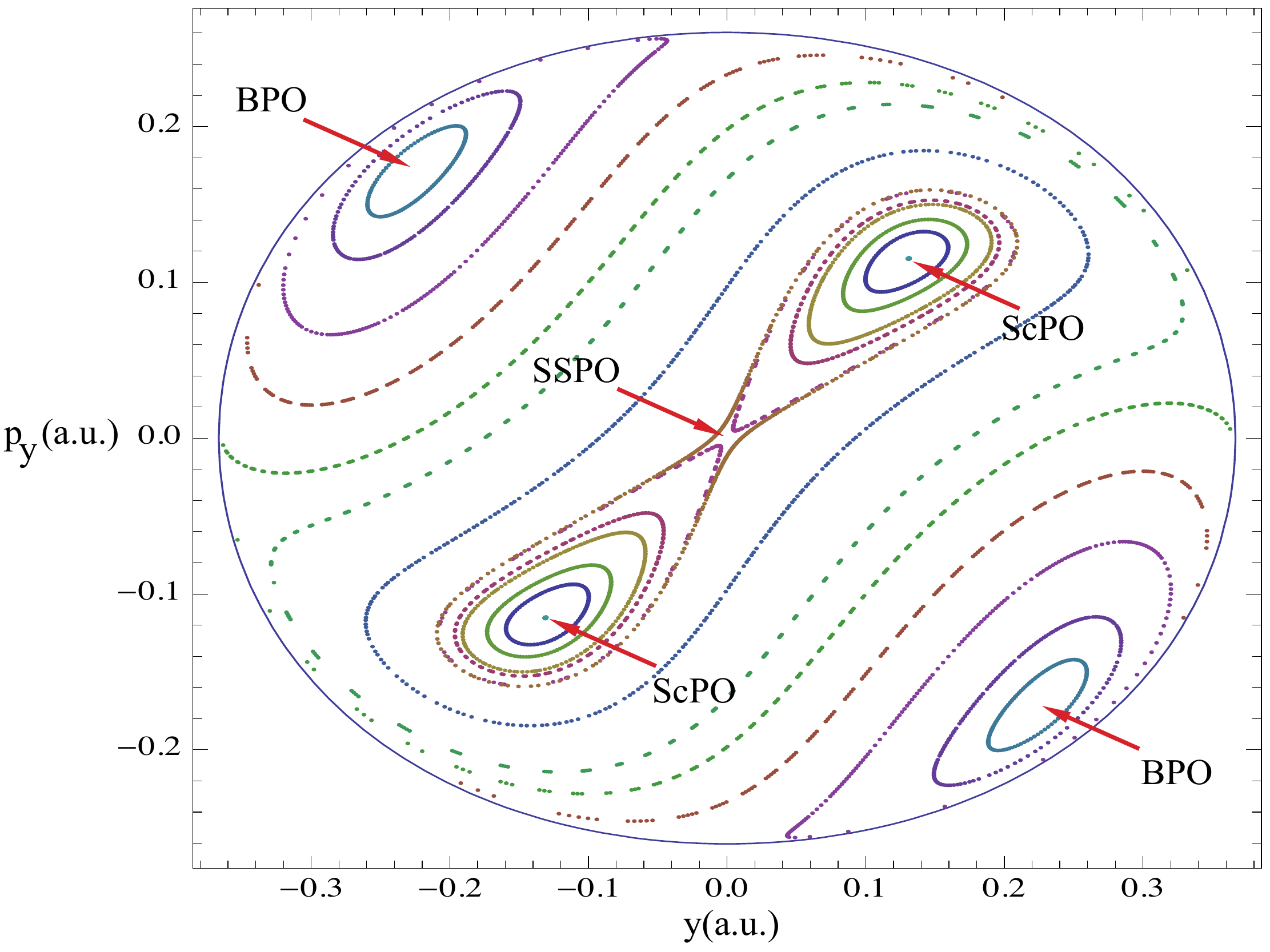}
 \caption{\label{fig:43}Poincar\'e surface of section $r = 3.40166\,$a.u. for dynamics within the centre manifold at the energy  $-4.3 \,$eV.}
 \end{figure}

In order to investigate the dynamics within the centre manifold in more detail, we choose a suitable Poincar\'e surface of section.  Since the centre manifold is four dimensional, the surface of section will have two dimensions and will be easy to visualize.  We pick the surface of section $y=0$ and use the canonically conjugate variables $r$ and $p_r$ as coordinates in the surface. The remaining momentum $p_y$ can be determined from the energy conservation condition
\begin{equation}
H(p_{r}, p_{y}, r, y=0) = E.
\label{Ham.cons}
\end{equation}
 We will always choose 
\[
p_{y}= p_{y}(p_{r}, r, E) > 0
\]
at the initial point, and in computing the Poincar\'e map we only consider intersections of a trajectory with the surface of section that have $p_y > 0$. The SSPO lies within the surface of section. Indeed, it bounds the area that is energetically accessible at a given energy. In contrast, the BPO appears as the central point in the low energy surface of section. For various energies, the surface of section is shown in Fig. \ref{fig:3}. At low energies, the intersections of a single trajectory with the surface of section lie on a closed curve, indicating quasi-periodic motion in accordance with the Kolmogorov--Arnold--Moser (KAM) theorem~\cite{KAM}. 

Some of the bifurcations of the fundamental periodic orbits, such as the loss and return of stability of the BPO, can also be seen in Fig. \ref{fig:3}. The bifurcation of the SSPO has an unusual appearance because the SSPO forms the boundary of the surface of section. As a consequence, the ScPO appears at the boundary and moves toward the centre of the surface of section. The Poincar\'e plots show only a single periodic point corresponding to the ScPO, as the chosen surface of section is $p_y>0$. A second periodic point is located in the surface $p_y<0$. Both periodic points can be seen in Fig.~\ref{fig:43}, which shows the Poincar\'e surface of section $r=r_\text{S}=3.40166\,$a.u. The SSPO intersects this surface transversely, and its bifurcations are therefore shown more clearly. The neighborhood of the SSPO in Fig.~\ref{fig:43} has the appearance one would expect close to a period doubling bifurcation. Note, however, that the situation is different from that shown in Fig.~\ref{fig:3}(e) in the neighbourhood of the BPO. As indicated by the colors, the two periodic points appearing there belong to two different periodic orbits, each of which has approximately the same period as the BPO. The two periodic points close to the SSPO in Fig.~\ref{fig:43} lie on a single periodic orbit of twice the period.

In addition to the fundamental periodic orbits, the surface of section plots show many other, longer periodic orbits that are not included in Fig. \ref{fig:4}. Of these there are, of course, infinitely many. Most important for our purposes is the observation that regions of chaotic dynamics appear and grow as the energy is increased. If the dynamics within the central sphere is chaotic, the central sphere might fail to be normally hyperbolic. We will investigate this question in the following section.

\section{\label{sec:level5}Breakdown of normal hyperbolicity}

For the hydrogen exchange reaction, we have seen that we can guarantee the existence of the central sphere for energies up to the conical intersection ridge without having to rely on its normal hyperbolicity. The full geometric structure of TST, however, also requires the existence of the reaction tubes, i.e., the stable and unstable manifolds of the central sphere. This can only be guaranteed if the central sphere is normally hyperbolic. We will now investigate the energy range in which this is the case.

We first need to state the condition of normal hyperbolicity more precisely. It is expressed in terms of Lyapunov exponents, which measure the rates at which nearby trajectories diverge under the dynamics: The Lyapunov exponent associated with variations of the initial conditions within the centre manifold ($\lambda_{int}$) should be less than the Lyapunov exponent away form the centre manifold ($\lambda_{ext}$). Thus normal hyperbolicity survives as long as $\lambda_{ext} > \lambda_{int}$. For  small enough energies above the saddle energy, the internal Lyapunov exponents  are zero because the dynamics within the central sphere is completely regular.

To compute a Lyapunov exponent for an arbitrary trajectory, consider a trajectory $x(t)$ and a neighboring trajectory $x(t)+\sigma(t)$. Both trajectories must satisfy the equations of motion~(\ref{eqmotion}). If the variation $\sigma$  is assumed to be infinitesimally small and the equations of motion are linearized in $\sigma$, we obtain the variational equations
\begin{equation}
\dot{\sigma} = -J \cdot  \mathcal{P} \cdot \sigma, \quad \sigma(t_0)=\sigma_0,
\label{eq3}
\end{equation}
where $\mathcal{P}$ is the Hessian matrix of the Hamiltonian,
\[
	\mathcal{P}_{ij}=\frac{\partial^2 H}{\partial x_i \partial x_j}.
\]
We integrate the combined systems (\ref{eqmotion}) and (\ref{eq3}) with arbitrary initial conditions $x_0$ and $\sigma_0$, and we ask how fast the length of the tangent vector $\sigma(t)$ will grow. The Lyapunov exponent of the trajectory starting at $x_0$ is defined by
\begin{equation}
\lambda (x_0, \sigma_0) = \lim_{t \rightarrow \infty} \frac{1}{t} \ln \frac{\| \sigma(t) \|}{\| \sigma_0 \|},
\label{LEgeneral}
\end{equation}
where $\| \sigma \|$ denotes the length of the vector $\sigma$. This definition corresponds to an exponential growth $\|\sigma(t)\|\propto e^{\lambda t}$. In general, the tangent vector $\sigma$ will quickly align itself with the direction in which the expansion rate is largest. The resulting Lyapunov exponent does then not depend on the arbitrarily chosen initial vector $\sigma_0$. There is an exception, however, for a trajectory in an invariant manifold: If the vector $\sigma$ is initially chosen tangent to the invariant manifold, it will remain tangent to it at all times. In this situation, we can meaningfully compute a Lyapunov exponent parallel to the invariant manifold and a Lyapunov exponent in the full phase space. The invariant manifold is normally hyperbolic if the latter is larger than the former.

The Lyapunov exponents are particularly easy to compute for a periodic orbit~\cite{Skokos01}. Because the evolution equation~\eqref{eq3} is linear in the variation vector $\sigma$, its solution can be written as $\sigma(t)=Y(t)\cdot \sigma_0$ with a matrix $Y(t)$ that does not depend on $\sigma$. The matrix $Y(T)$ is  called the monodromy matrix of the corresponding periodic orbit with period $T$, its eigenvalues $m_1, \dots, m_{2n}$ are the Floquet multipliers. For a periodic orbit with period $T$, we have $Y(\mu T)=(Y(T))^\mu$ for $\mu=1,2,\dots$. Therefore
\begin{equation}
\sigma(\mu T) = (Y(T))^{\mu} \cdot \sigma_0.
\label{Veqn}
\end{equation}
So, $m_1^\mu, ..., m_{2n}^{\mu}$ are the eigenvalues of $Y(\mu T)$. The spectrum of Lyapunov exponents of the particular periodic orbit is then 
\begin{equation}
\lambda_i = \lim_{\mu \rightarrow \infty} \frac{1}{\mu T} \ln | m_i^{\mu} | =  \frac{1}{ T} \ln | m_i | ,
\label{LEPO}
\end{equation}
and the largest of the Floquet multipliers $m_i$ will give the Lyapunov exponent~\eqref{LEgeneral}. For a periodic orbit in an invariant manifold we can use the eigenvectors of $Y(T)$ to distinguish whether eigenvalues correspond to variations parallel or transverse to the invariant manifold, and we can then choose the largest Lyapunov exponents in the parallel and transverse directions.

In a Hamiltonian dynamical system, the eigenvalues of the stability matrix $Y(T)$ will always occur in pairs $e^{\pm\lambda T}$ or $e^{\pm i\varphi T}$ with real numbers $\lambda$ and $\varphi$. These types of eigenvalues correspond to variations in unstable and marginally stable directions, and yield Lyapunov exponents $\lambda$ and $0$, respectively. A third possibility arises in Hamiltonian systems with three or more degrees of freedom: Eigenvalues can occur in quartets $e^{(\pm \lambda \pm i \varphi)T}$. This case is not relevant for our situation because the periodic orbits we are studying lie within the centre manifold, which is a subsystem with only two degrees of freedom.

The parallel and perpendicular Lyapunov exponents will in general be different for different trajectories in the invariant manifold, though they will be equal for trajectories on the same invariant torus or in the same chaotic sea. To verify normal hyperbolicity numerically, we must therefore calculate Lyapunov exponents for a large number of representative trajectories and check that the perpendicular Lyapunov exponent is larger than the parallel exponent in all cases.
  
\begin{table}
 \caption{Floquet multipliers of the Symmetric Stretch Periodic Orbit (SSPO).}
   \label{tab1}
\begin{ruledtabular}
 \centering
\begin{tabular}{ l|cc|cc}

 Energy & \multicolumn{2}{c|}{Within centre manifold} &  \multicolumn{2}{c}{Off centre manifold} \\ 
 \hline
-4.35             & \multicolumn{2}{c|}{$-0.949855 \pm 0.312691 i$} & 543.591 & 0.00183962 \\   
-4.32547 & \multicolumn{2}{c|}{ $-1 \pm  0.000746992 i$} & 513.284       & 0.00194824 \\ 
-4.02482      &-2.80253     &-0.35682    &2.80224           &0.356857 \\   
-4.02425     & -2.8049      & -0.356519   &  \multicolumn{2}{c}{$0.950972  \pm 0.309275  i$}  \\  
-4.02251        & -2.81218   &-0.355595    &  \multicolumn{2}{c}{$-0.98557 \pm 0.16925  i$}  \\ 
-4.02195     & -2.81447    &-0.355306    &   -2.83866   & -0.352278 \\ 
-4.0                & -2.9049      &-0.344246    & -52.8575     &-0.0189188 \\   
-3.5               & -4.63902   &-0.215563    & -1922.55     &-0.000520141 \\  
-3.0                &-5.8626       &-0.170573    &-6579.25      &-0.000151993 \\   
  \end{tabular}
  
\end{ruledtabular}
\end{table}

Earlier studies of the dynamics in the collinear subsystem\cite{Pollack79b,Pechukas82} found an energy interval in which the SSPO is stable against variations within that subsystem, which is transverse to the centre manifold. Also the narrow energy interval coincides with those values found recently by I\~narrea \textit{et al}\cite{Inarrea11} in the collinear case for the Porter-Karplus potential energy surface.   For these energies, the transverse Lyapunov exponent of the SSPO is zero, and it is clear that this situation must violate the condition of normal hyperbolicity as  soon as the transverse Lyapunov exponent decreases below that within the centre manifold. Table \ref{tab1} shows the Floquet multipliers of the SSPO within and transverse to the centre manifold. The SSPO is unstable within the centre manifold for those energies where it is stable in the collinear direction. As the energy increases further, the SSPO is unstable in both directions, but the instability in the collinear direction grows faster than that within the centre manifold, so that soon the SSPO does not violate the normal hyperbolicity of the central sphere any more.

On its own, this observation does not allow us to conclude that the centre manifold returns to being normally hyperbolic. It remains possible that  normal hyperbolicity could be broken by any orbit other than the SSPO. To check this, we have calculated the Lyapunov exponents for a variety of orbits in the centre manifold over a range of energies up to the conical intersection ridge. It turns out that across the entire range of energies no orbit apart from the SSPO violates normal hyperbolicity.

\begin{figure*}[t]
\includegraphics[width=6in]{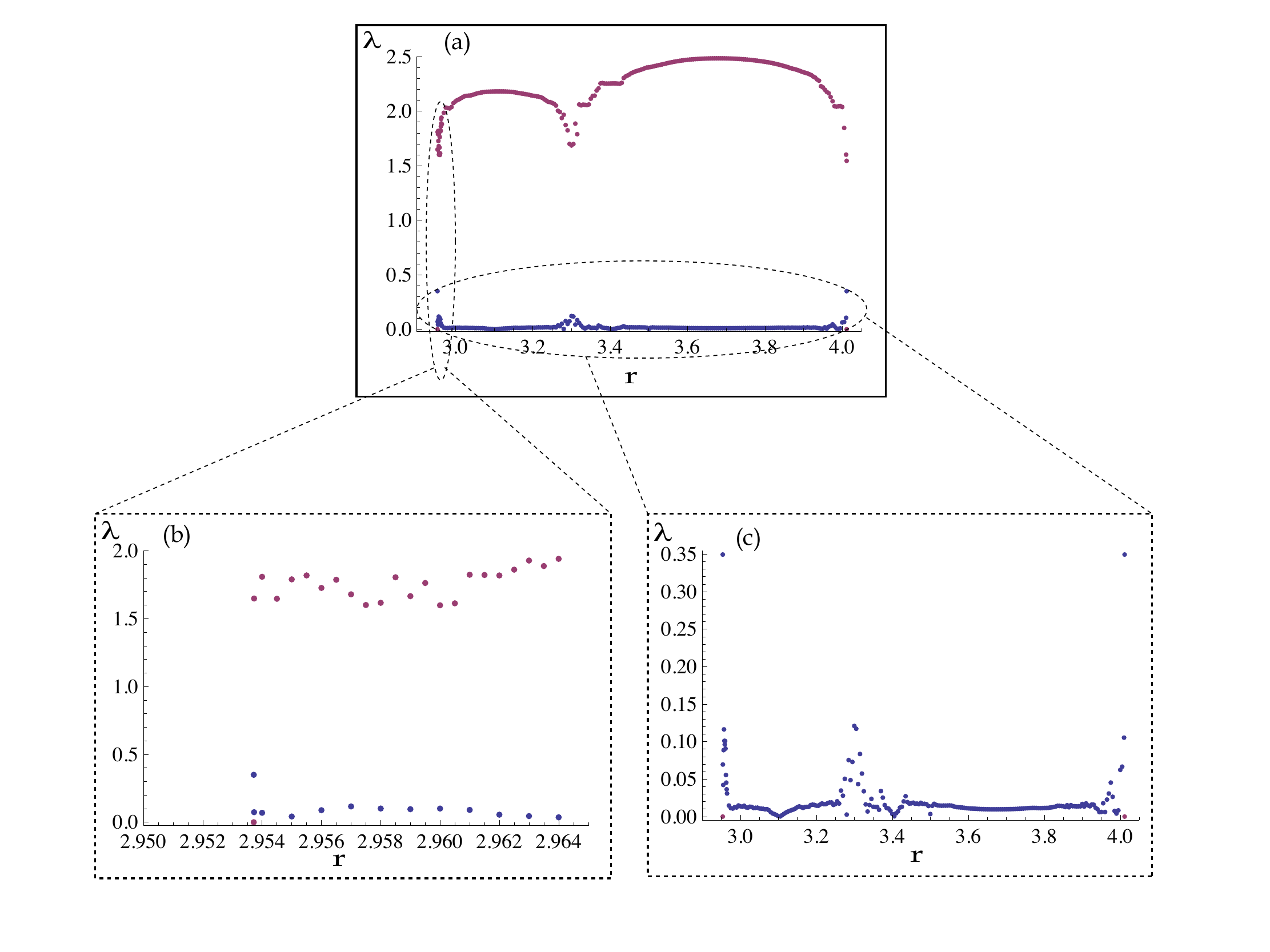}
 \caption{\label{fig:6}(Color online) The Lyapunov exponents within centre manifold (blue) and off centre manifold (red) through the section $p_r=0.0$ in the Poincar\'e surface for energy $E=-4.023$\,eV in the middle of the stable interval of SSPO.}
 \end{figure*}

As an example of these calculations, Fig.~\ref{fig:6}(a) shows the Lyapunov exponents within and off the centre manifold for the energy $-4.023$ eV, at which the SSPO is stable in the transverse direction, and for orbits on the line $p_r =0$ in the Poincar\'e surface of section. This section includes the SSPO, ScPO, BPO and both regular and chaotic nonperiodic orbits. Because the SSPO forms the boundary of the surface of section, the two points with the highest and lowest admissible values of $r$ correspond to the SSPO. The figure shows that normal hyperbolicity fails for these points, but not for any other orbits. The enlargement in Fig.~\ref{fig:6}(b) confirms this conclusion. Note that even for orbits arbitrarily close to the SSPO the transverse Lyapunov exponent is nonzero. Because the SSPO is unstable under variations within the centre manifold, an orbit that starts arbitrarily close to the SSPO will quickly move away from it, and its long term behaviour will be entirely different from that of the SSPO. For this reason, the Lyapunov exponents can be discontinuous at the SSPO. 

Fig.~\ref{fig:6}(c) focuses on the Lyapunov exponents within the centre manifold. They are much smaller than the transverse Lyapunov exponents, and the difference between trajectories on regular islands or in a chaotic sea can clearly be seen. For regular trajectories, we would expect these Lyapunov exponents to be zero. The numerical results show small, but finite values instead because the Lyapunov exponents were obtained by solving the equations of motion for a finite time only, whereas the definition~\eqref{LEgeneral} requires the limit of infinitely long simulation time. If the actual simulation time is increased, the resulting Lyapunov exponents become even smaller. 

We have so far focused only on the question whether the central sphere is normally hyperbolic, i.e., whether the ratio $k$ of the transverse to the parallel Lyapunov exponents is larger than one. In fact, the precise value of this ratio is also relevant because the fundamental theorems about normally hyperbolic invariant manifolds\cite{Fen71,Wigginsbook} guarantee that the stable and unstable manifolds of a NHIM exist and are differentiable at least~$k$ times. This result is important if normal form transformations are used to compute these manifolds, as they often have been\cite{Uzer02,Waalkens04b,Waalkens04,Li06,Li06a}. The normal form will effectively represent the invariant manifolds by Taylor series, which requires the existence of sufficiently high derivatives. Because derivatives of order higher than~$k$ are not known to exist, the use of high order normal forms is questionable if the ratio~$k$ is low. 

\begin{figure}[t]
\includegraphics[width=3.6in]{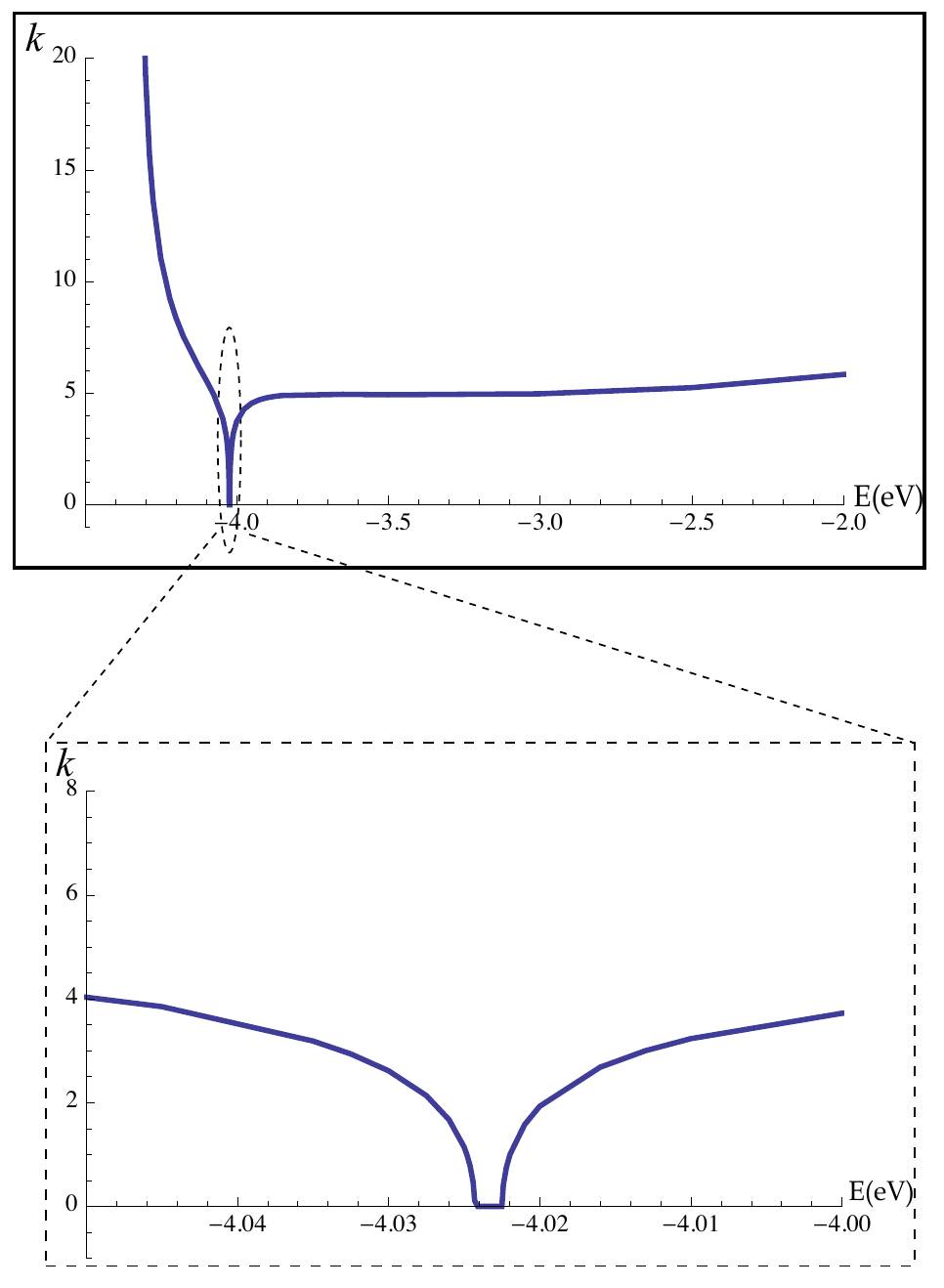}
 \caption{\label{fig:7}The ratio $k$ of two Lyapunov exponents within and off the centre manifold for SSPO.}
 \end{figure}

Fig.~\ref{fig:7} shows the ratio $k$ of Lyapunov exponents for the SSPO, which is the orbit that potentially violates normal hyperbolicity, for energies from the saddle point up to the conical intersection ridge. The ratio is infinite just above the saddle point because the Lyapunov exponent within the centre manifold is zero. It decreases from there and reaches zero when the SSPO is stable. It then rises again and reaches a nearly constant value of $k \approx 5$. As a consequence, we can expect the central sphere and its stable and unstable manifolds to be at least four times differentiable at all energies, except in a narrow range around the interval in which the SSPO is stable in the transverse direction.

\section{\label{sec:level6}Concluding remarks}

We have shown that in the hydrogen exchange reaction the central sphere exists for all energies below the conical intersection ridge and that it will possess stable and unstable manifolds for all energies outside a small interval. The surprising fact that the normal hyperbolicity of the central sphere is restored after it has been lost implies that the phase space structure fundamental to TST, which consists of the central sphere and its associated reaction tubes, will be in place even at energies high above the reaction threshold. Unfortunately, this result does not imply that the dynamics in the transition region will be simple. There will be homoclinic and heteroclinic tangles that lead to complex phase space geometry and consequently to complex dynamics. In the collinear subsystem of the full system, this complex behaviour has been shown by Davis\cite{davis87} and the most recently by  I\~narrea \textit{et al}\cite{Inarrea11}.

In the collinear subsystem, it is known that  dynamics is as simple as assumed by TST only if the PODS is unique\cite{Pollack79b}. However, additional periodic orbits arise at energies even lower than the energy at which the SSPO becomes stable, and  trajectories that violate the no-recrossing assumption of TST appear at the same energy.

Non-TST behavior in the full three-dimensional system must be at least as prevalent as in the two-dimensional subsystem. This means that even at energies at which the central sphere is normally hyperbolic, non-TST behavior must be present. These energies are both below and above the range in which normal hyperbolicity is broken.  Thus, while the results of the current paper demonstrate that the normal hyperbolicity of the central sphere is more robust than one might have anticipated, this robustness also implies that there is no direct link between the failure of TST and the violation of normal hyperbolicity. It now becomes a separate question to determine what dynamical effects, and what phase space structures,  cause the failure of TST. We will address this question in a forthcoming publication.

\begin{acknowledgments}
The first author is supported by Qassim University and the Ministry of Higher Education of Saudi Arabia. The research leading to these results has received funding from the People Programme (Marie Curie Actions) of the European UnionÕs Seventh Framework Programme FP7/2007-2013/ under REA grant agreement no. 294974.
\end{acknowledgments}

\appendix*
\section{The derivation of the kinetic energy expression}
\label{appendixA}

The kinetic energy is derived from that given, for example, by Waalkens et al \cite{Waalkens04}.  They study the HCN/CNH isomerization reaction in Jacobi coordinates: $r$ the distance between C and N, $R$ the distance between H and the centre of mass of CN and $\gamma$ the angle between H, the centre of mass of CN and C (i.e.: The atoms C, N and H  take the places of H1, H2 and H3, respectively, in our Fig.~\ref{fig:1}). The corresponding kinetic energy expression is
\begin{equation}
T = \frac{1}{2 \mu} p_{r}^{2} + \frac{1}{2 m} p_{R}^{2} + \frac{1}{2} \left(\frac{1}{\mu r^2} + \frac{1}{m R^2}\right) p_{\gamma}^{2},
\end{equation}
where $\mu = m_C m_N / (m_C +m_N)$ is the reduced mass of CN and $m = m_H (m_C + m_N) / (m_H + m_C + m_N)$ is the reduced mass of the full system.
In the exchange hydrogen reaction, we have three identical atoms. Thus $\mu$ and $m$ become $\frac{1}{2} m_H$ and $\frac{2}{3} m_H$, respectively. As a result the kinetic energy has the form
\begin{equation}
T= \frac{1}{m_H} p_{r}^{2} + \frac{3}{4 m_H} p_{R}^{2} + \left(\frac{1}{m_H r^2} + \frac{3}{4 m_H R^2}\right) p_{\gamma}^{2}.
\label{prekinetic}
\end{equation}
It is singular when $R=0$. This is the case for symmetric collinear configurations such as the saddle point that is of central importance in our study. To avoid this singularity, we replace the polar coordinates $R$ and $\gamma$ by Cartesian coordinates $x$ and $y$, as shown in Fig.~\ref{fig:1}. The coordinate systems are related by
\begin{align*}
x &= R \cos \gamma, \qquad &y &= R \sin \gamma, \\
R^2 &=x^2+y^2, \qquad & \gamma &= \arctan\left(\frac{y}{x}\right).
\end{align*}
We use $r$ as the third coordinate as before.

The generating function $W$ associated with this transformation is \[
W = p_r r + p_x R \cos \gamma + p_y R \sin \gamma.
\]
It yields the following transformation of momenta:
\begin{eqnarray*}
p_R  =  \frac{\partial W}{\partial R} & = & p_x \cos \gamma +  p_y \sin \gamma \\
	& = & \frac{xp_x+yp_y}{R}, \\
p_{\gamma} = \frac{\partial W}{\partial \gamma} & = &-p_xR \sin \gamma +  p_y R \cos \gamma \\
	& = & -yp_x + x p_y.
\end{eqnarray*}
Substituting these results into (\ref{prekinetic}), we get
\[
T = \frac{1}{m_{H}}\left[ p_{r}^{2} + \frac{3}{4} (p_{x}^{2}+p_{y}^{2}) + \frac{(x p_y - y p_x)^2}{r^{2}} \right],
\]
which is the result used in~\eqref{newHamil}.



%

\end{document}